\documentclass{workshop}
\usepackage{cite}
\usepackage[dvipsnames]{xcolor}
\begin{document}

\title{
Reverse Shocks in Short Gamma-Ray Bursts \\ 
{\large -- The case of GRB 160821B and prospects as gravitational-wave counterparts --} 
}

\author{
Gavin P Lamb$^1$ 
\\[12pt]  
%
$^1$  School of Physics and Astronomy, University of Leicester, University Road, Leicester, LE1 7RH, UK \\
%
\textit{E-mail: gpl6@leicester.ac.uk} 
}

\abst{
The shock system that produces the afterglow to GRBs consists of a forward- and a reverse-shock. 
For short GRBs, observational evidence for a reverse-shock has been sparse -- however, the afterglow to GRB\,160821B requires a reverse-shock at early times to explain the radio observations. 
GRB\,160821B is additionally accompanied by the best sampled macronova without a gravitational-wave detection, and an interesting late time X-ray afterglow behaviour indicative of a refreshed-shock. 
The presence of an observed reverse-shock in an on-axis short GRB means that the reverse-shock should be considered as a potential counterpart to gravitational-wave detected mergers. 
As a gravitational-wave counterpart, the afterglow to an off-axis GRB jet can reveal the jet structure -- a reverse-shock will exist in these structured jet systems and the signature of these reverse-shocks, if observed, can indicate the degree of magnetisation in the outflow. 
Here we show the case of GRB\,160821B, and how a reverse-shock will appear for an off-axis observer to a structured jet.
}

\kword{Yamada conference LXXI: proceedings --- gamma-ray bursts: GRB 160821B, general --- gravitational wave: electromagnetic counterparts}

\maketitle
\thispagestyle{empty}

\section{Introduction}

For a relativistic shell expanding into a medium, two shocks will form; 
a forward shock that propagates into the external medium, and a reverse shock that propagates into the shell \citep{sari1995}.
In describing the reverse shock, two regimes are usually discussed, these are the thin shell, or Newtonian shock case, and the thick shell, or relativistic shock case \citep[etc]{kobayashi2000}.
The relativistic shell will decelerate at the reverse shock crossing time in both regimes \citep{kobayashi1999}.
The duration and energetics of short gamma-ray bursts (GRBs) imply that any reverse shock in these systems will typically be described by the thin shell case \citep[2019]{lamb2018a}.

Emission from a reverse shock is an important probe of the conditions in a GRB outflow towards the central engine.
Modelling the observed afterglow emission enables constraints to be placed on the magnetisation and the bulk Lorentz factor $\Gamma_0$, where $F_{max, r} = \Gamma_0 F_{max, f} C_F R_{B}$, and here $F_{max}$ is the maximum synchrotron flux from the reverse and forward shock, $r$ and $f$ respectively, $C_F$ is a correction factor, see \citet{harrison2013}, and $R_{B} \equiv \varepsilon_{B,r}/\varepsilon_{B,f}$ is the magnetisation parameter, see \citet{zhang2003}.

The reverse shock in short GRBs has been notoriously difficult to observe \citep{lloydronning2018}, however, the short GRB\,051221A has evidence\footnote{GPL thanks Alexander van der Horst for kindly pointing out this paper in a comment following the conference talk} of a reverse shock with a radio frequency detection and upper-limits \citep{soderberg2006}, and recently reverse shock emission has been shown to be consistent with the optical afterglow to the candidate short GRB\,180418A \citep{becerra2019}, and radio observations of the afterglow to the short GRB\,160821B require a reverse shock component \citep{lamb2019a, troja2019}.
These successes in observing the reverse shock in short GRBs, although rare, raise the prospect of identifying the reverse shock emission in gravitational-wave detected neutron star mergers, where the systems jet-axis is likely misaligned for an observer and so the afterglow and GRB will be observed off-axis \citep[2018, 2019]{lamb2017}.

\begin{figure*}[t]
\centering
\includegraphics[width=\textwidth]{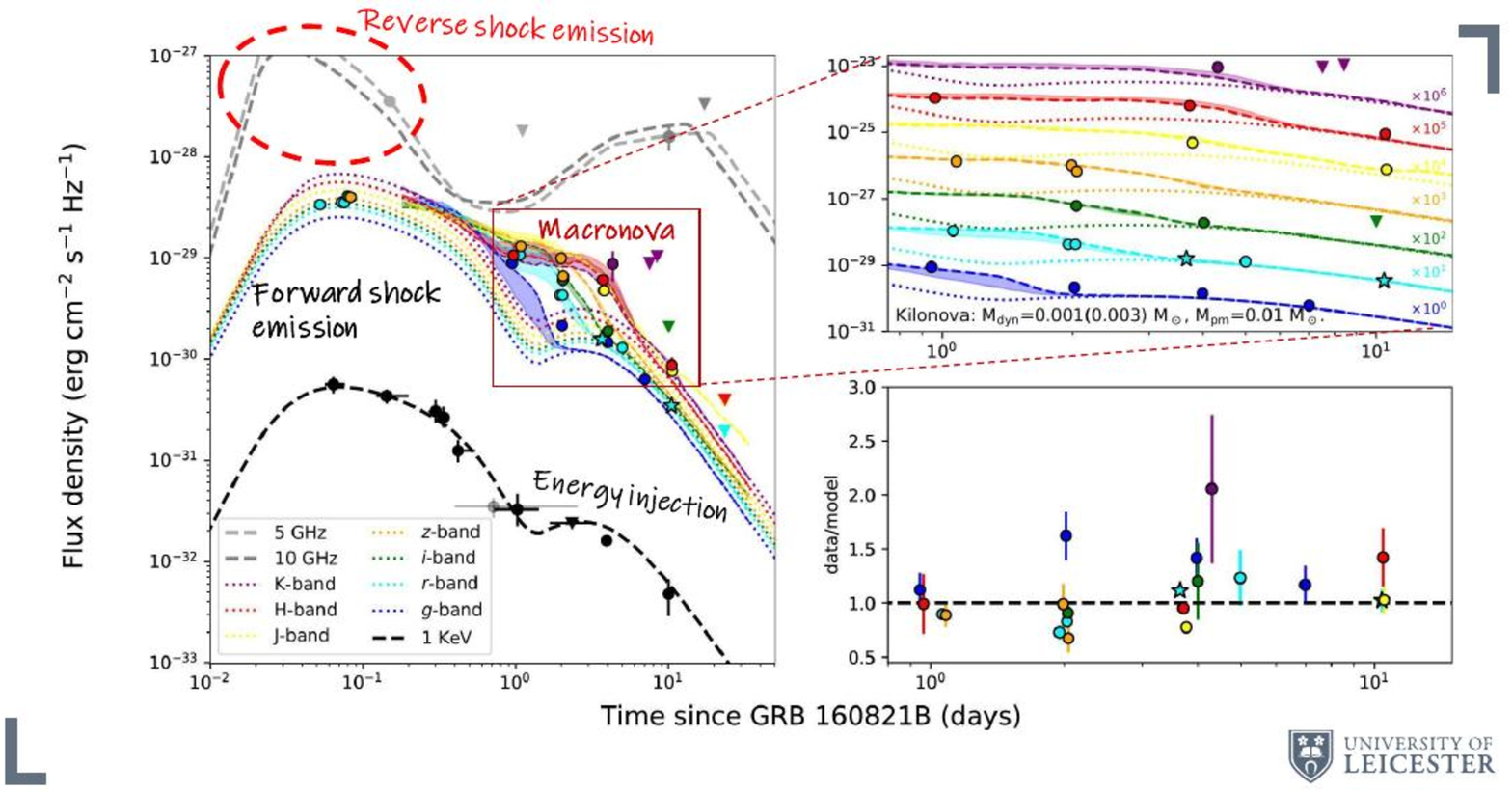}
\caption{The broadband afterglow to GRB\,160821B. Dashed or dotted lines are the model light-curves, filled circles are data, and triangles upper-limits, see \citet{lamb2019a} for details. Left: the afterglow from radio to X-ray frequencies (top to bottom), a reverse shock contribution is required at early times to explain the radio observation, while at late times energy injection is needed to explain the X-ray, optical and radio data. From $\sim1$--$5$ days an excess at optical and infrared indicates a macronova. Right top: expands the macronova and afterglow model with the data. Right bottom: shows the model fits versus data residual.}
\label{fig:first}
\end{figure*}

\section{GRB 160821B}


The short GRB\,160821B, at a redshift of $z=0.162$, had an isotropic gamma-ray energy of $E_{\gamma, iso}=(2.1\pm0.2)\times10^{50}$\,erg, based on the 8--10,000\,keV band fluence of $(2.52\pm0.19)\times10^{-6}$\,erg\,cm$^{-2}$ \citep{lu2017}.
X-ray, optical, and radio frequency observations of the afterglow to GRB\,160821B were performed from 0.06--23.23\,days after the burst;
for a full list of the observations used here see \citet{lamb2019a}.
Early infrared observations put limits on the presence of a macronova \citep{kasliwal2017}, however, more complete broadband observations revealed emission at optical and infrared frequencies in excess of that expected from afterglow modelling of the X-ray and radio data \citep{lamb2019a, troja2019}.

In Fig. \ref{fig:first} we show the afterglow data and our preferred model.
The complex behaviour of the afterglow is explained variously by:
the contribution of a reverse shock travelling into a mildly magnetised ($R_B\sim8$) shell at early times ($\sim0.1$\,days), we estimate a bulk Lorentz factor for the initial outflow $\Gamma_0\sim60$;
a jet break at $\sim0.3-0.4$\,days followed by an injection of energy from a fallback powered second jet episode\citep{rosswog2007, kagawa2019}\footnote{This second jet episode is slower than the initial GRB creating jet and likely responsible for the X-ray extended emission lasting $\sim300$s following the main burst} that re-brightens the afterglow \citep[etc]{granot2003};
and a macronova contribution at $\sim1-5$\,days.
The best-fitting macronova model \citep{kawaguchi2018} consists of two-components:
a dynamical ejecta mass of $\sim0.001$\,M$_\odot$, and a post-merger or secular ejecta mass of $\sim0.01$\,M$_\odot$.

\section{Reverse Shocks as Gravitational Wave Counterparts}

\begin{figure*}[t]
\centering
\includegraphics[width=\textwidth]{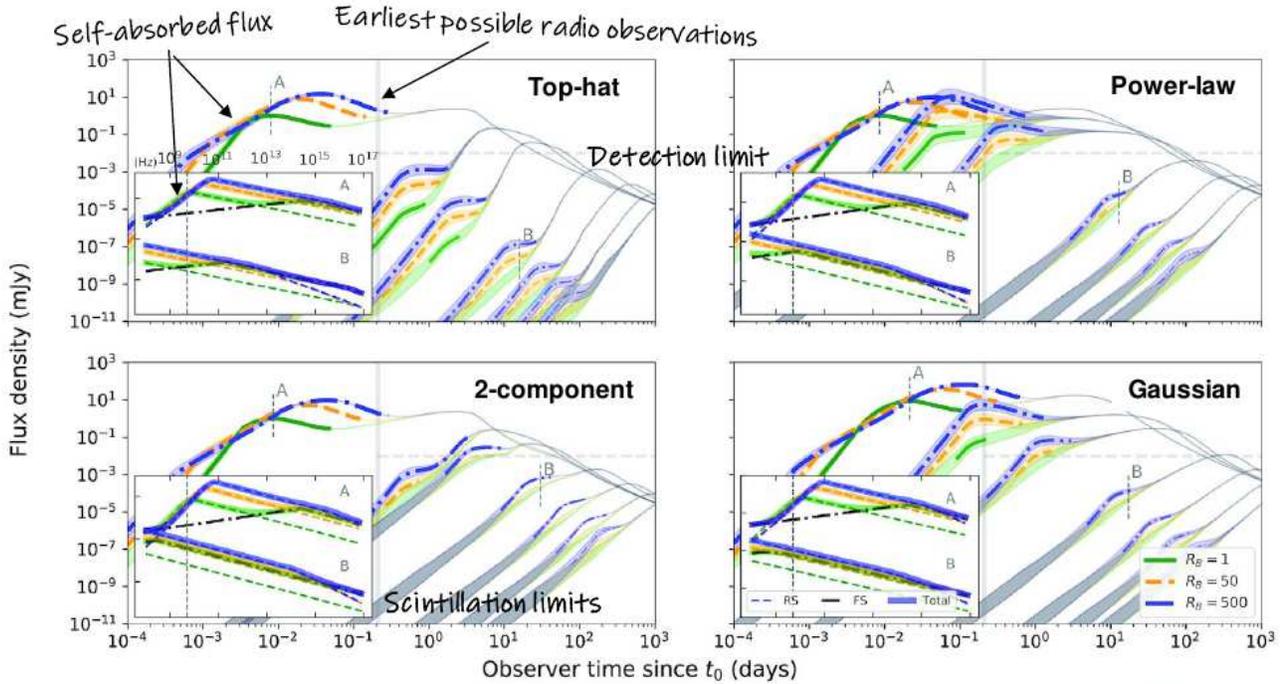}
\caption{The afterglow light-curve for four jet structure models following those in \citet{lamb2017} but including sideways expansion, synchrotron self-absorption, and a reverse shock for an ejecta shell characterised by a magnetic parameter $R_B=[1,50,500]$, bold lines in green (solid), yellow (dashed), and blue (dash-dotted) respectively, see \citet{lamb2019}. The light-curves for an afterglow at 5\,GHz and 100\,Mpc are shown at  $[0^\circ, 12^\circ, 18^\circ, 36^\circ, 54^\circ, 72^\circ,]$ and $[90^\circ]$. The effects of scintillation are shown with the grey shaded regions at early times while the size emitting region is still small \citep{granot2014}. Insets on each panel show the spectra at the times indicated by `A' and `B'.}
\label{fig:second}
\end{figure*}

The GRB\,170817A and its macronova and afterglow in association with the gravitational-wave detected merger of a binary neutron star at $\sim40$\,Mpc \citep{abbott2017} has shown that short GRBs are produced in binary neutron star mergers.
The late-time afterglow to GRB\,170817A indicated that the resultant jets from neutron star mergers, when viewed at a higher inclination than the central jet axis, will reveal the outflow structure \citep{lamb2017}.
The post-peak rapid decline of the late-time afterglow \citep{lamb2019b, troja2019b}, along with VLBI observations \citep{ghirlanda2019, mooley2018}, cleared any ambiguity in the jet-core dominated origin of the afterglow \citep{lamb2018b}.

For neutron star mergers discovered via gravitational-waves, the reverse shock in the afterglow can potentially be observed for systems that are inclined $<20^\circ$, at a luminosity distance $<100$\,Mpc, and that have some lateral structure; see Fig \ref{fig:second} and \citep{lamb2019}.
Additionally, a magnetisation parameter $R_B\sim$\,a few is required and observations should ideally commence at $\sim0.1$\,days post merger and at radio frequencies $<100$\,GHz.
For such cases, scintillation may complicate the observations, however, carefully measured scintillation can be used to constrain the size of the emitting region \citep{granot2014}.

\label{last}

\end{document}